\documentclass[aps,prd,10pt,amsmath,amsfonts,twocolumn,nofootinbib]{revtex4-1}
\usepackage[T1]{fontenc}

\usepackage{float,amsmath, amssymb,epsfig,graphicx,mathrsfs}
\usepackage{amsfonts,bbm,color,bm}
\usepackage{amsmath}
\usepackage[colorlinks]{hyperref}
\usepackage[figure,table]{hypcap}
\usepackage[textsize=tiny]{todonotes}
\usepackage{subfigure}

\hypersetup{
	bookmarksnumbered,
	pdfstartview={FitH},
	citecolor={green},
	linkcolor={darkblue},
	urlcolor={darkblue},
	pdfpagemode={UseOutlines}}
\definecolor{darkgreen}{RGB}{20,100,20}
\definecolor{darkblue}{RGB}{0,0,130}
\definecolor{darkred}{rgb}{.8,0,0}
\newcommand{\psii}{\psi_\text{I}}
\newcommand{\psiii}{\psi_\text{II}}
\newcommand{\phii}{\phi_\text{I}}
\newcommand{\phiii}{\phi_\text{II}}
\newcommand{\ali}{\alpha_\text{I}}
\newcommand{\alii}{\alpha_\text{II}}
\newcommand{\bei}{\beta_\text{I}}
\newcommand{\beii}{\beta_\text{II}}

\def\I{ \mathbbm{1} }
\def\A{\cal A}

\begin{document}

\title{Modes mismatch induced variation of quantum coherence for two-mode localized Gaussian states in accelerated frame}

\author{Xiaolong Gong$^1$}
\author{Yue Fang$^{2\dag}$}
\author{Tonghua Liu$^{1,3\star}$}
\author{Shuo Cao$^{3,4}$}

\email{liutongh@yangtzeu.edu.cn;caoshuo@bnu.edu.cn}
\email{fangyue@hbmzu.edu.cn;}
\affiliation{1. School of Physics and Optoelectronic, Yangtze University, Jingzhou 434023, China; \\
2. College of Intelligent Systems Science and Engineering, Hubei Minzu University, Enshi 445000, China;\\
3. Institute for Frontiers in Astronomy and Astrophysics, Beijing Normal University, Beijing 102206, China; \\
4. Department of Astronomy, Beijing Normal University, Beijing 100875, China;}

\begin{abstract}
Quantum coherence is the basic concept of superposition of quantum states and plays an important role in quantum metrology. We show how a pair of uniformly accelerated observers with a local two-mode Gaussian quantum state affects the Gaussian quantum coherence. We find that the quantum coherence decreases with increasing acceleration, which is due to the Unruh effect that destroys the quantum resource. Essentially, the variation of quantum coherence is caused by the modes mismatch between the input and output mode. Through 2000 randomly generated states, we demonstrate that such mismatch is dominated by the acceleration effect and mildly affected by the waveform parameters. Moreover, the squeezing parameter acted as a suppressor of the reduced coherence, but it tended to be invalid in the high squeezing. In addition, the squeezing parameter can act as a suppressor of the reduced coherence, but the effect of the squeezing parameter tends to be ineffective under high squeezing conditions.
\end{abstract}

\maketitle

\section{Introduction}
The coherent superposition of quantum states is one of the decisive characteristics that distinguishes quantum mechanics from the classical field \cite{1980PThPS..69...80L}. Quantum coherence can be used to explain some intriguing and fantastic phenomena in quantum optics \cite{1963PhRv..131.2766G,1991PhRvL..67.1855S,1994JMOp...41.2467A}. Particularly, quantum coherence could exist in the single systems, which is different from the properties of the quantum entanglement and quantum discord. Based on this unique feature, quantum coherence has great applications in quantum communication and quantum information processing \cite{Divincenzo2000, 1997PhRvL..79..289B,2014PhRvL.113n0401B,2014PhRvL.113q0401G,2016arXiv160502458C,PhysRevA.93.032326,2017RvMP...89d1003S,liu3}. For example, Wu \textit{et al.} have explored how to obtain the maximal coherence and analyzed the wave-particle duality relation based on the coherence measurements \cite{2017Optic...4..454W}. Winter and Yang have established an operational theory of quantum coherence by focusing on the optimal rate of performance of certain tasks \cite{Winter2016}. Furthermore, a series of efforts have been made to quantify quantum coherence \cite{Yuan2015,Yu2016,Marvian2016,Zhang2016,Chen2016,Tan2017}.

Due to the real world is always accompanied by gravity and relativistic effects, the preparation and transmission of quantum states cannot be accomplished without a gravitational field. With the development of quantum technology, more and more attention has been paid to the performance of traditional quantum tasks and quantum resources under the influence of relativistic effects, and quantum coherence is no exception.  Chen \textit{et al.} have showed that the relative entropy of coherence is destroyed as increasing acceleration of the
detectors \cite{Chen2016a}. The freezing condition of coherence for accelerated free modes in a relativistic setting beyond the single-mode approximation was studied by \cite{Huang2018}. Moreover, the dynamics of the quantum coherence has been given in \cite{Wang2016}. In addition, some attempts have previously been made to perform experimental tests of quantum resources in relativistic coordinate systems, with the realization of photonic entanglement in an accelerated setting \cite{2017NatCo...815304F}. However, it is well known that one cannot get the full information from a global mode in the experiment, since the method of single mode approximation is delocalized in practice (single mode approximation method involves the Unruh mode \cite{Peres2004}). Such issue has been extensively discussed in several recent works \cite{liu1,liu2,liu4}, taking gravitational or accelerated field as lossy channel rather than global free models, and furthermore calculating the fidelity of the lossy channels \cite{2022QuIP...21..397L}.

Recently, an ideal scheme \cite{Dragan2013, Dragan2013a} was proposed to solve this problem, named \textit{localized mode}. By storing the squeezed Gaussian states in a simulated cavity capable of transport quantum states, they successfully realized the localization of Gaussian states. With the proposal of the localized Gaussian state, the performance of various resources in such framework became one of the most concerned topics. In 2013, Dragan \textit{et al.} proposed a localized projective operator formalism and extracted the vacuum entanglement by a pair of counter-accelerating detectors \cite{Dragan2013}. Later, Ahmadi \textit{et al.} continued the research and found the sudden death of entanglement due to the spatial separation between the observers \cite{Ahmadi2016}. Grochowski \textit{et al.} reported an entanglement degradation due to the mismatch of the two observers's accelerations \cite{2017PhRvD..95j5005G}. Moreover, Fang \textit{et al.} found that a higher Gaussian interferometric power can be obtained from the resource state with larger entanglement \cite{2019QuIP...18..248F}.
In addition, D\c{e}bski and  Dragan studied relative purity with two proper accelerated observers and  initial level of squeezing  of the symmetric Gaussian state  numerically. These results showed that the purity of
pure state is reduced with acceleration and the effect is stronger for the states
with the larger entanglement \cite{2018PhRvD..98b5003D}.

In this work, we following the previous work \cite{2018PhRvD..98b5003D}, and will investigate the performance of Gaussian quantum coherence with localized two-mode Gaussian quantum states in a relativistic setting. We choose a two-mode squeezed vacuum state of the localized wave packet as input state, and transform it to the accelerated frame of reference corresponding to a pair of uniformly accelerated observers. We study how the detector's proper acceleration affects Gaussian quantum coherence. Moreover, through a simple quantification of the mode mismatch, we exhibit the variation of the mismatch versus various parameters and discuss its influence on the coherence. Our paper is organized as follows: In Sec.\ref{Sec2}, we review the Gaussian quantum channel and the selection of the modes. In Sec.\ref{Sec3}, we introduce the Gaussian quantum coherence and analyze its variation with the accelerations. The discussion about the mode mismatch will be presented in Sec.\ref{Sec4}. Finally, Sec.\ref{Sec5} provides conclusions. In this paper, we use units such that $c = \hbar = k_B = 1$.

\section{The Framework}\label{Sec2}
In our scheme, we consider two uniformly accelerating observers, Alice and Bob, in an accelerated frame of reference which transform a two-mode state prepared in an inertial frame from initial frame to the observer's. The localization process of quantum states is equivalent to a noisy Gaussian channel, which can be characterized by explicit analytic expressions with no approximations.  In this section, we will briefly reintroduce the noisy quantum Gaussian channel from \cite{Ahmadi2016}, and a detailed introduction could be found in Section II there.

\subsection{The Gaussian channel}
We investigate a real scalar quantum field $\hat\Phi$ with a mass $m$ in $1+1$-dimensional Minkowski spacetime, and the evolution of $\Phi$ is governed by the Klein-Gordon equation $(\square+m^2)\hat{\Phi} = 0$, where $\square=\eta^{\mu\nu}\partial_\mu\partial_\nu$. The implicit Klein-Gordan scalar product can be written as:
\begin{align}\label{KG}
	(\phi_1,\phi_2) = i\int_{\Sigma}\mbox{d}x \left(\phi_1^\star \partial_t\phi_2-\phi_2\partial_t\phi_1^\star \right),
\end{align}
where $\Sigma$ is a spacelike Cauchy surface and the imaginary factor guarantees that $(\phi_1,\phi_2) = (\phi_2,\phi_1)^\star = -(\phi^\star_2,\phi^\star_1)$ \cite{Birrell1984}.

One of the most important innovations in the localized quantum state framework is the introduction of modified Rindler coordinate, which lifts the restriction of geometry of the Rindler chart and enables observers to have arbitrary proper accelerations and the minimal distance between them. The corresponding transformation from Minkowski coordinates to the modified Rindler coordinates system were:
\begin{align}\label{coor_transformation}
t &= \pm \chi \sinh a \eta \\ \nonumber
x &= \pm \chi \cosh a \eta \pm \frac{D}{2},
\end{align}
where $x$ and $t$ are Minkowski coordinates, $\chi$ and $\eta$ are Rindler coordinates, and $a$ is an arbitrary positive parameter (interpreted as the proper acceleration of a trajectory). In our scheme, we are interested in analyzing the situation that the distance $D = 0$ between two Rindler wedges. It means that the two Rindler wedges have a common apex.

A total of four localized wavepackets were taken in consider, where $\phi_n$ \footnote{Here we remind the reader that n can be used to label any other modes of the orthogonal basis. But these remaining modes are useless to us, the specific reasons will be mentioned below. So, for convenience, in this paper we stipulate that $n \in \lbrace \text{I},\text{II} \rbrace$, which corresponding to the Rindler region $\text{I}$ or region $\text{II}$.} are assumed to be inertial, with associated annihilation operators $\hat{f}_n$, and the mode $\phi_n$ is localized within the corresponding region at $t=0$. In the accelerated frame of reference, we consider two accelerated modes $\psi_{\text{I}}$ and $\psi_{\text{II}}$ that always remain localized within the corresponding region. In addition, $\phi_n$ and $\psi_n$ should only contain the positive frequency part \cite{Ahmadi2016}. By assuming these two sets of modes are orthogonal, hence:
\begin{align}\label{relation}
(\phi_\text{I},\phi_\text{II}^\star)&=0,   [f_\text{I},f_\text{II}^\dag]=0,\\ \nonumber
(\psi_\text{I},\psi_\text{II}^\star)&=0,   [d_\text{I},d_\text{II}^\dag]=0,
\end{align}
and the field operator $\Phi$ can be decomposed by these two sets of modes with respect to the Klein-Gordon inner product:
\begin{align}\label{decompose}
\hat{\Phi}=\sum_n[\phi_n\hat{f}_n+H.c.] = \sum_n[\psi_n\hat{d}_n+H.c.].
\end{align}

Our goal is perform such Gaussian channel to transform the state of  $\phi_{\text{I}}$ and $\phi_{\text{II}}$ into the state of $\psi_{\text{I}}$ and $\psi_{\text{II}}$. To do so, we need to perform a Bogolyubov transformation from the inertial to the uniformly accelerated frame, and tracing out the modes with $n \notin \lbrace \text{I},\text{II} \rbrace$. A general form of the quantum channel to describe such transformation can be written as:
\begin{subequations}\label{channel}
	\begin{align}
	\vec{X}^{(d)} &=  M \vec{X}^{(f)},\label{channel1}\\
	\sigma^{(d)} & =  M\sigma^{(f)} M^T+N,
	\label{channel2}
	\end{align}
\end{subequations}
where $\vec{X}^{(f)}$ and $\vec{X}^{(d)}$ are the first moments corresponding to the inertial and accelerated modes, $\sigma^{(f)}$ and $\sigma^{(d)}$ known as the covariance matrix of the inertial wavepackets' state and observers' state, respectively. Due to our resource state will be a two-mode state, $M$ and $N=N^T$ are $4\times 4$ real matrices. The channel is uniquely characterized by the specification of $M$, which is also a symplectic matrix \cite{Simon1988}, and the matrix $N$ denotes the noise in the quantum channel. Moreover, this channel can completely retain Gaussianness, i.e., the resulting output state will be Gaussian if the input resource state is Gaussian.

\subsection{Input and output modes}
As discussed in \cite{Doukas2013, Dragan2013, Dragan2013a, Ahmadi2016}, the wave packets are approximately localized and consist of only positive frequency in respective rest frames. The input modes consists of a Gaussian envelope and sinusoidal modulation, as:
\begin{align}\label{inputmode}
\phi_n(x,0)&=\pm C \exp[-2(\frac{x_0}{L}\log{\frac{x}{x_0}})^2]f(x), \nonumber \\
{\partial}_t \phi_n(x,0)&=-i  \Omega_0\phi_n(x,0),
\end{align}
where $f(x)=\sin\left(\sqrt{\Omega_0^2-m^2} (x-x_0)\right)$, $x_0$ is the central position of the mode function, $L$ denotes its width, $C$ is a normalization constant,  and the upper (lower) sign $\pm$ refers to $n=\text{I (II)}$. The frequency $\Omega_0$, about which the spectrum of the mode function is centered, has to be sufficiently large to effectively damp the negative frequencies, i.e. satisfying $\Omega_0\gg\frac{1}{L}$. Additionally, we introduce a zero-frequency cutoff to completely eliminate the distribution of negative-frequency part.
\begin{figure}[ht!]
	\begin{center}
		\includegraphics[height=1.8in,width=2.8in]{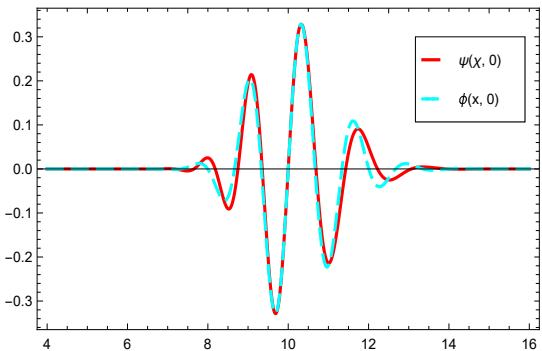}
		\caption{The shapes of input and output modes. We fixed ${\A}_\text{I}={\A}_\text{II}=0.1, L=2$, $\Omega_{0}\approx 5$, and the field mass $m=0.1$.\label{fig1}}
	\end{center}
\end{figure}
The input modes do not directly determine the output modes $\psi_n$, however, which can be obtained by replacing the trigonometric functions with modified Bessel functions and substituting the Rindler coordinates in place of inertial ones. The output mode functions are given by
\begin{align}
\label{outputmode}
&\psi_n(\chi,0)=C' \exp[{-2\left( \frac{x_0}{L}\log\frac{\chi}{x_0}\right)^2}]g(\chi), \nonumber \\
&{\partial}_\tau \psi_n(\chi,0)=\mp i  \Omega_0\psi_n(\chi,0),
\end{align}
where $C'$ is the normalization constant and the upper (lower) sign $\mp$ corresponds to $n=\text{I (II)}$. The modulating function of the output mode is replaced by the combination of modified Bessel functions of the first kind $I_{iv}$, with $g(\chi)=\Im \left[I_{-i\frac{\Omega_0}{{\A}_n}}(m|x_0|)I_{i\frac{\Omega_0}{{\A}_n}}(m|\chi|)\right]$, and $|x_0| = \frac{1}{{ \A}_n}$. Furthermore, the proper acceleration ${\A}_n$ satisfies the boundary condition $1/{ \A}_n\gg L$ due to the fact that the accelerating mode functions $\psi_n$ are far from the event horizon compared to their size $L$.

In Fig.~\ref{fig1}, the shapes of the localized wave packets were presented, and the parameters were $L=2, \Omega_0\approx 5, {\A}_\text{I}={\A}_\text{II}=0.1$ and the field mass $m=0.1$. It is clearly that there is a tiny but inevitable mismatch between the input and output modes, i.e., the two modes can't overlap completely. Although the Gaussian waveform is one of the optimal shapes against the effects of mode mismatch \cite{Rohde2005}, the introduction of relativistic effects can still exacerbate this effect. This is because one of the modes is stationary while the other is accelerating.

\begin{figure}[ht!]
	\begin{center}
		\includegraphics[height=3.2in,width=3.2in]{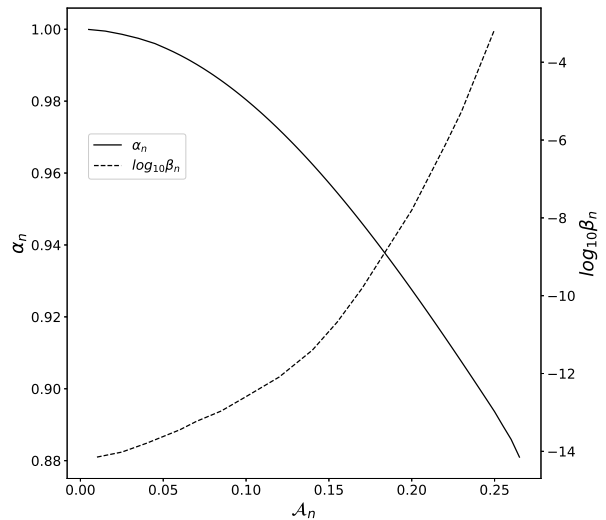}
		\caption{The $\alpha_n$ and $\beta_n$ versus ${\A}_n$. We fixed $L=2$, $\Omega_{0}\approx 5$ and the field mass $m=0.1$. \label{fig2}}
	\end{center}
\end{figure}
The $\alpha_n$ and $\beta_n$ are defined as $\ali=(\psii,\phii)$, $\bei=-(\psii,\phii^\star)$, $\alii=(\psiii,\phiii) $ and $\beii=-(\psiii,\phiii^\star)$. Once we determine the mode functions, the values of $\alpha_n$ and $\beta_n$ could be written as a function of the acceleration ${\A}_n$, and we plot this relation in Fig.~\ref{fig2}. One can see that $\alpha_n$ decreases monotonically with an increasing ${\A}_n$. The matrix $M$ could be written as
\begin{widetext}
	\begin{equation}\label{M}
	M=
	\left(\begin{matrix}
	\text{Re}(\ali-\bei) & -\text{Im}(\ali+\bei) & 0 & 0 \\
	\text{Im}(\ali-\bei) & \text{Re}(\ali+\bei) & 0 & 0 \\
	0 & 0 & \text{Re}(\alii-\beii) & -\text{Im}(\alii+\beii) \\
	0 & 0 & \text{Im}(\alii-\beii) & \text{Re}(\alii+\beii)
	\end{matrix}\right).
	\end{equation}
\end{widetext}
Due to the fact that the value of $\beta_n$ is lower by several orders of magnitude than that of $\alpha_n$, we can omit $\beta_n$ and simplify the matrix $M$ to
\begin{align}
M=\alpha_\text{I}\I\oplus\alpha_\text{II}\I,
\end{align}
where $\I$ is a $2\times 2$ identity matrix. The specific form of the matrix $N$ is relatively complex, and we refer the readers to \cite{Ahmadi2016} for details. We only give the simplified form of the matrix $N$, which is
\begin{align}
N=(1-\alpha_\text{I}^2)\I\oplus(1-\alpha_\text{II}^2)\I.
\end{align}

\section{Coherence}\label{Sec3}
In this section we briefly review the measurement of quantum coherence for a general two-mode Gaussian state, and calculate the effect of relativistic acceleration through the knowing $\alpha_n$. It's well known that any two-mode Gaussian state can be fully described by a covariance matrix \cite{Simon2000,Duan2000}, which in a block form is
\begin{align}
\sigma=\left(
\begin{array}{cccc}
a & 0 & c_1 & 0\\
0 & a & 0 & -c_2\\
c_1 & 0 & b & 0\\
0 & -c_2 & 0 & b\\
\end{array}
\right).
\end{align}
The symplectic eigenvalues of the $\sigma$ are $2\nu_{\mp}^2=\Delta\mp\sqrt{\Delta^2-4\det\sigma}$, where $\Delta=b^2+d^2+2c_1c_2$. According to \cite{Xu2016}, the coherence measure $C(\sigma)$ was given in terms of the displacement vectors and covariance matrix. For convenience, we set the displacement vector to be zero, and the analytical expression of the $C(\sigma)$ of a Gaussian state could be written as
\begin{align}
	C(\sigma)=&-S(\sigma)+\Sigma_{i=1}^{2}[(\bar{n}_{i}+1) \log _{2}(\bar{n}_{i}+1)\\\nonumber&-\bar{n}_{i} \log _{2} \bar{n}_{i}],
\end{align}
where the mean occupation values are
\begin{align}
	\bar{n}_{1}=\frac{1}{4}(\sigma_{11}+\sigma_{22})\\
	\bar{n}_{2}=\frac{1}{4}(\sigma_{33}+\sigma_{44})\nonumber
\end{align}
and the von Neumann entropy is
\begin{align}
	S(\sigma) &=f(\nu_-)+f(\nu_+)\\\nonumber
	 &=\frac{\nu_-+1}{2} \log _{2} \frac{\nu_-+1}{2}-\frac{\nu_--1}{2} \log _{2} \frac{\nu_--1}{2}\\\nonumber
	 &+\frac{\nu_++1}{2} \log _{2} \frac{\nu_++1}{2}-\frac{\nu_+-1}{2} \log _{2} \frac{\nu_+-1}{2}.
\end{align}

\begin{figure}[!htbp]
\centering
\includegraphics[scale=0.5]{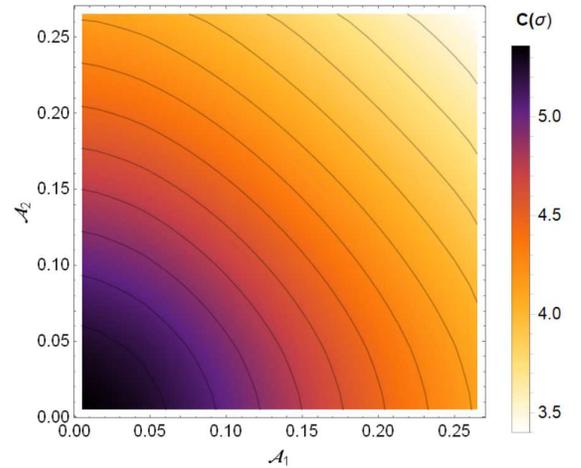}
\caption{The coherence $C(\sigma)$ of the two-mode squeezed vacuum resource state as detected by asymmetrically accelerating Alice and Bob. The original state is characterized by a squeezing coefficient $r=1$.}
\label{fig3}
\end{figure}

We begin with a two-mode squeezed vacuum state that characterized only by the squeezing parameter $r$ as input state, and the covariance matrix is
\begin{align}
\label{input}
\sigma^{(f)}&=\left(\begin{array}{cccc}
\cosh 2r  &0 &\sinh 2r &0\\
0  &\cosh 2r &0 &-\sinh 2r\\
\sinh 2r &0  &\cosh 2r &0\\
0 &-\sinh 2r &0 &\cosh 2r
\end{array}\right).
\end{align}
Applying the Eq.~\eqref{input} to the Eq.~\eqref{channel2}, the covariance matrix $\sigma^{(d)}$ observed by two accelerating observers is given by
\begin{align}\label{outputstate}
\sigma^{(d)}&=\left(\begin{array}{cccc}
X  &0 &Y &0\\
0  &X &0 &-Y\\
Y &0  &Z &0\\
0 &-Y &0 &Z
\end{array}\right)
\end{align}
where
\begin{eqnarray}
X&=&\ali^2 \cosh{(2r)}-\ali^2+1,\nonumber\\
Y&=&\ali\alii\sinh{(2r)},\nonumber\\
Z&=&\alii^2 \cosh{(2r)}-\alii^2+1.
\end{eqnarray}
Assuming that two observers moving with arbitrary accelerations, the variation of the coherence is plotted in Fig.~\ref{fig3}.  When the accelerated parameters $\A_\text{I}$ or $\A_\text{II}$ reach to zero in the Fig.3, the values of coherence recover the initial value in the inertial frame. Furthermore, the quantum coherences of number values are not of particular concern, since the normalized factor are not used in here. Our work focuses more on the trend between quantum coherence and squeezing parameters, and so on, which is more directly physical. Fixing one of the accelerations (e.g., $\A_\text{I}$), we can see that $C(\sigma)$ monotonically decreases with the other increasing acceleration. In fact, with the increasing of accelerations, both the mode mismatch and Unruh noise are strengthened. However, the Unruh noise in such scheme is only play a minor role \cite{Dragan2013a}, and we will discuss the effects of mode mismatch in the next section. This result indicate that, when performing more reliable quantum information tasks, the relativistic effects cannot be ignored.

\section{The Effects of Mode Mismatch}\label{Sec4}
\begin{figure}[!htbp]
	\centering
	\includegraphics[scale=0.5]{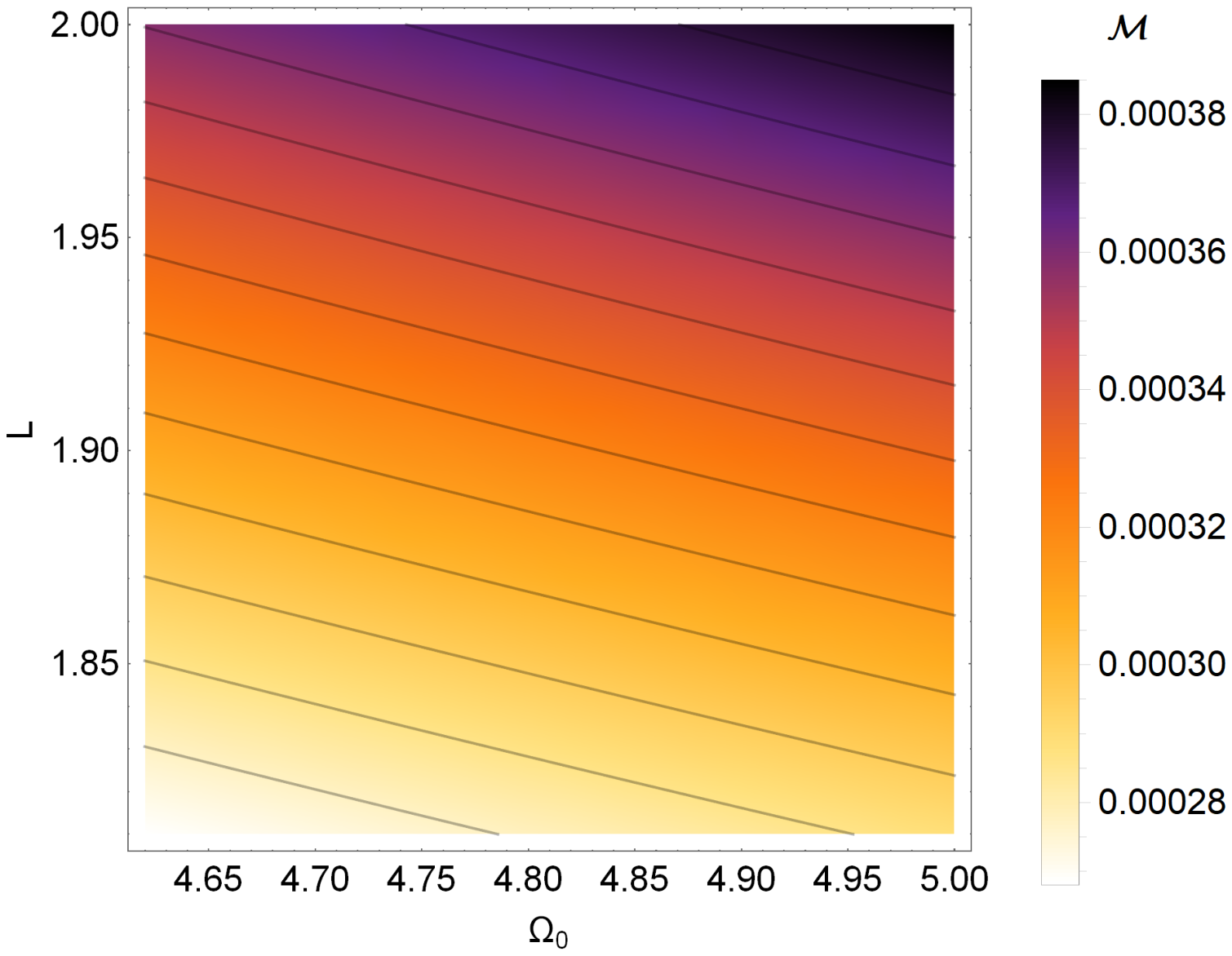}
	\includegraphics[scale=0.5]{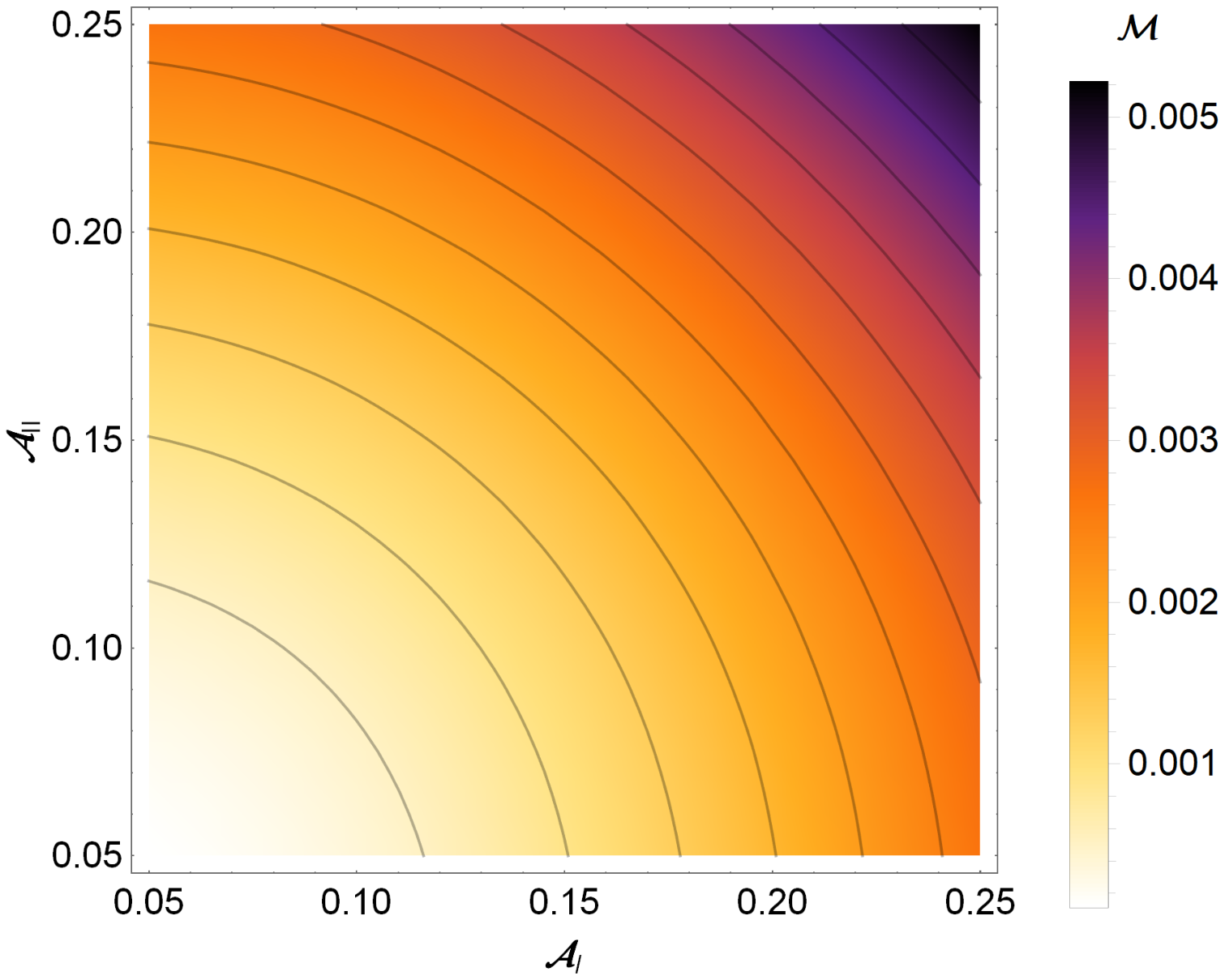}
	\caption{The variation of the mode mismatch $\cal M$. The top and bottom panels give the case that fixes the accelerations (${\A}_\text{I}={\A}_\text{II}=0.1$) and the waveform parameters ($\Omega_{0}=4.7$ and $L=2$), respectively. The field mass $m$ is chosen to be $0.1$.}\label{fig4}
\end{figure}

The mode mismatch always occurs when the origin mode is different from the target mode, and it will be caused by the beam displacement, tilting, or beam size difference in the experiment \cite{Roh2021}. Different from the case of classical, the loss of squeezed light due to the mode mismatch can not be simply compensated by means of increasing the optical power \cite{Toeyrae2017}. Furthermore, the mode mismatch is an intrinsic property between our input and output modes and cannot be removed by any amendments to the protocol \cite{Dragan2013a,Ahmadi2016}.  Unfortunately, there is no authoritative quantification of the mode mismatch yet. In order to visualize the impacts of the mismatch. In spite of  the parameter $\alpha$ seems like a natural candidate for quantification of the mode mismatch, it is a non-analytic quantity related to acceleration, which itself needs to be solved by numerical method, so it is difficult to achieve quantization for mismatch of modes. Thus, it is necessary to analyze the impact of such mismatch. We mathematically define the mismatch $\cal M$ as the shape difference between the input and output modes, i.e.,
\begin{align}
	{\cal M} = \frac{\Sigma_{i=1}^k(\phi_{n}(x_i,0)-\psi_{n}(\chi_i,0))^2}{k},
\end{align}
where $x_i$ and $\chi_i$ are the Minkowski and Rinder coordinates with a step of 0.01 in $[0.02, 1/A+3L]$, respectively. Noticed that the shapes of the two modes will also change with the waveform parameters ($L$ and $\Omega_{0}$), but this effect is negligible compared to that caused by the acceleration. By fixing the acceleration and waveform parameters separately, the numerically calculated $\cal M$ was shown in Fig.~\ref{fig4}. It is obvious that the influence of acceleration on mode mismatch is much greater than that of waveform parameters, both in terms of amount and variation amplitude. Alternatively, the mode mismatch is dominated by the effect of relativistic acceleration. Moreover, it also could explain why the entanglement degradation which led by the acceleration effect is much stronger than that led by the waveform parameters (see Fig. 5 and Fig. 6 in \cite{Ahmadi2016}).

\begin{figure}[!htbp]
	\centering
	\includegraphics[scale=0.37]{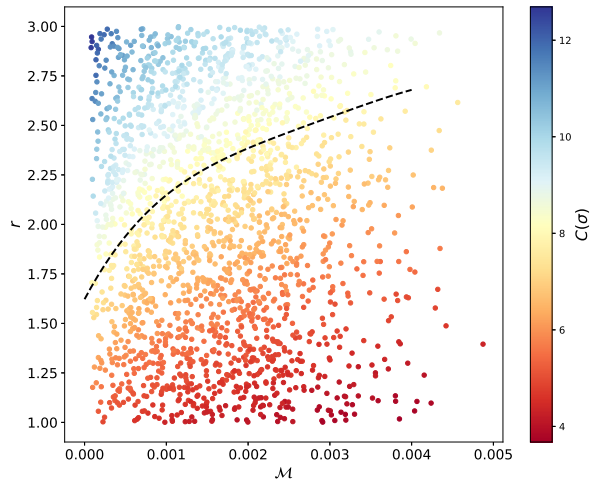}
	\caption{The coherence $C(\sigma)$ of 2000 randomly generated states versus the squeezing parameter $r$ and mode mismatch $\cal M$, and the field mass $m$ is chosen to be $0.1$. The dashed line indicates the contour line corresponding to the median of the obtained $C(\sigma)$.}\label{fig5}
\end{figure}

In order to visualize the impact of the mismatch on $C(\sigma)$, 2000 states with random waveform parameters and accelerations were generated, and the resource state was characterized by the squeezing parameter $r\in[1,3]$. As shown in Fig.~\ref{fig5}, the numerically computed quantum coherence $C(\sigma)$ was presented. The amount of $C(\sigma)$ decreases monotonically with the increasing mode mismatch, but the squeezing operation suppresses the coherent decreasing caused by the mode mismatch. In fact, several quantum resources would be degraded or even destroyed by the relativistic effects \cite{Dragan2013a,Ahmadi2016,Grochowski2017,2019QuIP...18..248F}, and would be protected by the squeezing operation. However, with the squeezing parameter increases, such suppression effect becomes weaker and even tends to be invalid, which is indicated by the dashed contour line in Fig.~\ref{fig5}. In the previous studies of the quantum entanglement, the suppression effect was demonstrated to become stronger with the increases of the squeezing parameter \cite{Grochowski2017}, which is opposite to our case. This gives us
the guidance to choose appropriate physical parameters to perform more reliable quantum information task by using such as quantum coherence resources with the inevitable relativistic effects.

\section{Conclusions}\label{Sec5}
In this work, we considered two uniformly accelerating observers, Alice and Bob, which transformed a two-mode squeezed vacuum state from the initial frame to the observers' frame, which was regarded as a Gaussian quantum channel.  By neglecting $\beta$, a simplified form of the channel was obtained and it could be used to easily compute the output state $\sigma^{(d)}$, and the coherence that could be extracted from the input two-mode squeezed vacuum state was analyzed.

We showed that quantum coherence monotonically decreases with increasing observers ( Alice and Bob) of acceleration. And we affirmed the reason of decreased coherence with the increasing relativistic acceleration, essentially due to the growing mismatch between the inertial and accelerating modes. Through a definition of the mode mismatch, we demonstrated that the mode mismatch was dominated by the acceleration effect, although it was also affected by waveform parameters. This result indicate that, when performing more reliable quantum information tasks, the relativistic effects cannot be ignored. Moreover, the decreasing of the coherence would be suppressed by the squeezing operation. However, such suppression effect tend to be invalid with the squeezing parameter increases, which is opposite to the case in the entanglement. In the case of unavoidable relativistic effects, this provides guidance for us to select appropriate physical parameters for the preparation of quantum coherence to perform more reliable quantum information tasks.

As the final remark, although the influence of Unruh effect on quantum coherence cannot be detected by current technology, it on quantum  coherence or other quantum resources
cannot be ignored, especially when performing relativistic
quantum information tasks. According to the equivalence principle, the effects of acceleration are equivalence with the effects of gravity, our results could be in principle applied to dynamics of
 quantum coherence between the Earth and satellites.

\textbf{Data Availability Statement}
My manuscript has no associated data.

\textbf{Competing Interests Statement}
The authors have no competing interests to declare that are relevant to the content of this article.

\section*{Acknowledgments}
We thank Anonymous editors for their constructive suggestions which have greatly improved the quality of our paper.
Liu. T.-H was supported by National Natural Science Foundation of China under Grant No. 12203009; Chutian Scholars Program in Hubei Province.

\twocolumngrid
\bibliography{reference}
\end{document}